\documentclass{optica-article}
\journal{oe}
\articletype{Research Article}
\usepackage{caption}
\usepackage{lipsum}
\usepackage{lineno}

\begin{document}
\title{Bipartite communication game-based empirical demonstration of non-trivial and universal non-contextuality without unphysical idealizations}
\author{Xuan Fan,\authormark{1} Ya Xiao,\authormark{1,2} and Yongjian Gu\authormark{1,3}}

\address{\authormark{1}College of Physics and Optoelectronic Engineering, Ocean University of China, Qingdao City, Shandong Province, People's Republic of China, 266100\\}

\email{\authormark{2}xiaoya@ouc.edu.cn\\
\authormark{3}yjgu@ouc.edu.cn}
\begin{abstract}
Universal contextuality is the leading notion of non-classicality even for single systems, showing its advantage as a more general quantum correlation than Bell non-locality, as well as preparation contextuality. However, a loophole-free experimental demonstration of universal contextuality at least requires that both operational inequivalence and compatibility loopholes are closed, which have never been simultaneously achieved to date. 
In our work, we experimentally test universal contextuality through (3,3) and (4,3) communication games, simultaneously restoring operational equivalence and circumventing the compatibility loophole. 
Our result exhibits the violation of universal non-contextuality bound by 97 standard deviations in (3,3) scenario, and 107 deviations in (4,3) scenario.  Notably there are states which exhibit locality but reveal universal contextuality in both two scenarios. In addition, our result shows that universal contextuality is more general than preparation contextuality in (3,3) scenario, while equivalent to preparation contextuality in (4,3) scenario.
\end{abstract}

\section{\label{Intro}Introduction}
Two famous no-go theorems have shone since 1960s like a pair of twin stars, and have left afterward enlightening paths in both a deepened understanding of quantum mechanical nature and an expanded vision of rangy applications. One of them is the prestigious Bell’s theorem \cite{bell1964einstein}, whose vitality has been well proven in self-testing \cite{mayers2004self, coladangelo2017all}, certification of randomness \cite{pironio2010random, acin2012randomness, acin2016optimal}, and device-independent cryptography \cite{ekert1991quantum, acin2007device, barrett2013memory}, to name just a few. Another one is known as Kochen-Specker theorem \cite{bell1966problem, kochen1967problem}, which is advantageous for its testability even in a single system with dimension $d\geqslant3$, and is now seen of broad applications in fault-tolerant quantum computation \cite{raussendorf2013contextuality,howard2014contextuality, delfosse2015wigner, delfosse2017equivalence, bermejo2017contextuality, mansfield2018quantum, emeriau2022quantum}, increasing channel capacity \cite{cubitt2010improving, cubitt2011zero} and parity-oblivious multiplexing \cite{spekkens2009preparation, chailloux2016optimal, hameedi2017communication, ambainis2019parity, saha2019state}, etc.\par
However, since its origin, the traditional version of non-contextuality has been limited by several fatal shortcomings. First, it is not suitable for an arbitrary physical theory but is limited to quantum theory. Second, it applies only to sharp measurements, excluding other experiment procedures such as unsharp measurements. Last but not least, it applies only to deterministic hidden-variable models and not to other models in quantum theory, where the outcomes of measurements are only probabilistically determined from the investigated system. To address those limitations, Spekkens proposed a generalized version of the original definition based on operational theory \cite{spekkens2005contextuality}. In this model, Spekkens extends the formulation to preparation, transformation, and measurement non-contextuality. Notably, when a system satisfies both preparation and measurement non-contextual preconditions in the definition, it is considered universal non-contextual. This generalization expands and deepens the notion of quantum contextuality as an important quantum correlation, removing the limited scope of applicability and establishing it as an irreplaceable resource in quantum information processing tasks.\par

Just like its counterpart Bell non-locality, the demonstration of contextuality is vital for understanding its nature, and requires eliminating additional assumptions by closing “loopholes”. Among those loopholes, two of the greatest importance are known as operational inequivalence and compatibility loophole, which have never been closed simultaneously in any experiment or proposal up to date \cite{cabello1998proposed, simon2000feasible, cabello2008proposed, klyachko2008simple, cabello2008experimentally, badziag2009universality, michler2000experiments, moussa2010testing, lapkiewicz2011experimental,hasegawa2006quantum, bartosik2009experimental, amselem2009state}. Operational equivalence, as a necessary precondition for testing universal contextuality \cite{spekkens2005contextuality}, has failed to be achieved in most works mentioned above. The only exception proposes a method that may solve the problem of operational inequivalence \cite{mazurek2016experimental}, but is only practiced in heralded single photon system, and thus limited in other possible applications. What’s more, their test of universal contextuality overlooks compatibility loophole, which, despite discussed in various experiments by different means \cite{kirchmair2009state, huang2013experimental, zhang2013state, borges2014quantum, xiao2018experimental, hu2016experimental}, has never been realized when testing universal contextuality. Thus, it would be a significant achievement for us to close those two loopholes at the same time.

In this work, we present an experimental demonstration of universal contextuality free of both operational inequivalence and compatibility loopholes. First, we introduce a generalized (m,n) bipartite communication game (with m preparations and n measurements) that has been used to study different physical models, especially for disproving local and non-contextual hidden variable theories \cite{hameedi2017communication, pan2019revealing}. Second, we carry out experiments by sharing a series of entangled states based on the optical system, satisfying both operational equivalence and no-signaling condition as two irreplaceable preconditions. 
Our experimental results indicate that universal contextuality is violated respectively by 97 and 107 standard deviations, respectively, in (3,3) and (4,3) scenarios.
Moreover, this work clearly reveals how different quantum correlations are deeply connected with each other. We demonstrate how the violation recedes by decreasing the entanglement of shared states. Notably, in this process, the disappearance of non-locality comes much earlier than universal contextuality, leaving a region which carries both universal contextuality and Bell locality, namely showing non-trivial contextual behaviours. Finally, we discuss some open questions that warrant further exploration, which could open a new page in the study of contextuality through strictly loophole-free experiments in the future.

\section{\label{Theory}Theoretical framework}
\subsection{\label{Games}Communication games as tests of universal contextuality}

  To show how universal contextuality can be revealed through communication games, we consider a generalised bipartite Bell scenario \cite{pan2019revealing} which involves $m$ preparations and $n$ measurements, i.e., $(m,n)$ scenario. Suppose the preparer and the measurer initially share a two-particle state $\rho_{AB}$. In this communication game, the preparer proceeds by choosing a uniformly random input $x\in \{1,2,...,m\}$, performing the corresponding preparation $P_x\in \{P_1,P_2,...,P_m\} $ and recording the output $a_x\in \{-1,1\} $. Similarly, the measurer performs the corresponding measurement $M_y\in \lbrace M_1,M_2,...,M_n \rbrace $ according to his received input $y\in \{1,2,...,n\}$, and records the output $b_y\in \{-1,1\} $. After each round of measurement, the preparer and the measurer communicate through a classical channel. The rules for winning this game are as follows:
For $x+y=$max$[m,n]+1$, where max[] is to return the largest item in the square bracket, the outputs should satisfy $a_x \neq b_y$; And if $x+y \neq$max$[m,n]+1$, then $a_{x} = b_{y}$. Thus the success probability can be written as
\begin{equation}
\begin{aligned}
\mathbb{P}_{m,n}&=\frac{1}{mn}[\sum_{x=1}^m\sum_{y=1}^n(p(a_x\neq b_y|P_{x},M_{y};x+y={\rm max}[m,n]+1))\\
&+(p(a_x=b_y|P_{x},M_{y};x+y \neq {\rm max}[m,n]+1))],  
\end{aligned}
\label{uc}
\end{equation}
where $p(a_x,b_y|P_x,M_y)$ denotes the probability  when the preparer implements a preparation $ P_x$ with outcome $a_x$ and the measurer performs a measurement $M_y$ with outcome $b_y$. It's worth mentioning that the communication game in $(n,m)$ scenario can easily be converted to the $(m,n)$ case, by a simple role-exchanging between the preparer's  preparations and  the measurer's measurements.
Thus, Eq. (\ref{uc}) can be used to test the existence of universal contextuality. And it can also be rewritten with a suitable Bell-like parameter $ \beta_{mn} $ as follows:

\begin{equation} 
\begin{aligned}
\mathbb{P}_{m,n}=\frac{1}{2}+\frac{\beta_{mn}}{2mn},   
\end{aligned}
\label{UC}
\end{equation}
where $\beta_{mn}=\sum_{x=1}^m\sum_{y=1}^n\langle P_{x} M_{y};x+y={\rm max}[m,n]+1)\rangle - \sum_{x=1}^m\sum_{y=1}^n\langle P_{x} M_{y};x+y\neq{\rm max}[m,n]+1)\rangle $,
$\langle P_{x} M_{y}\rangle=\sum_{a_{x},b_y} (-1)^{ a_x b_y } p(a_x,b_y|P_x,M_y)$. It has been demonstrated that, in bipartite Bell scenario, the universal non-contextuality bound $\mathbb{P}_{mn}^{NUC}$ is no higher than the local realist bound $\mathbb{P}_{mn}^{L}$.
Therefore, with appropriate states and measurements, non-classicality can be revealed through the violation of universal non-contextuality, even if quantum theory doesn’t violate the local realist bound. It’s worth noting that the smaller $m$ and $n$ are, the easier it is to demonstrate universal contextuality. In our experiment, we mainly demonstrate universal contextuality by taking the (3,3) and (4,3) scenarios as two examples due to the following reasons. First, they take the simplest setting in our model, and are more suitable for a proof-of-principle demonstration of universal contextuality like this; Second, despite its simple form, from Eq.(\ref{UC}) we can tell that they have the largest quantum value, and thus the experiment would be more convincing for a greater violation against non-contextual bound, as will be shown later; Last but not least, those two scenarios have seen a wide range of application in many quantum information tasks, especially for (4, 3) scenario, which has been proved to be deeply related to 3→1 quantum random access code (QRAC)\cite{spekkens2009preparation}. 

\subsection{\label{Condition}Operational Inequivalence and Compatibility Loophole}
A loophole-free demonstration of universal contextuality requires at least two necessary preconditions: the restoration of operational equivalence, and the closure of compatibility loophole.
In communication games, different preparation operations $\{P_1,P_2,...,P_m\} $ are considered as operationally equivalent if, after normalization, they yield the same reduced quantum state over outcomes for an tomographically complete set of measurements, i.e.,
\begin{equation}
\begin{aligned}
\begin{split}
\sum_{a_{1}} p(a_{1} \vert P_{1})\rho(a_{1} \vert P_{1})&=\sum_{a_{2}} p(a_{2}  \vert P_{2})  \rho(a_{2} \vert P_{2})\\&=....=\sum_{a_{m}} p(a_{m} \vert P_{m})\rho(a_{m} \vert P_{m}),
\end{split}
\label{op1}
\end{aligned}
\end{equation}
where $p(a_x \vert P_x)$ and $\rho(a_x \vert P_x)$ respectively denote the probability and the conditional state of the preparer when she implements a preparation $ P_x$ with outcome $a_x$ \cite{pusey2018robust}.
Similarly, the measurements $\{M_1,M_2,...,M_n\} $, are said to be operationally equivalent if they yield the same reduced quantum state after normalization over outcomes for a tomographically complete set of preparations, i.e.,
\begin{equation}
\begin{aligned}
\begin{split}
\sum_{b_{1}} p(b_{1} \vert M_{1})\rho(b_{1} \vert M_{1})&=\sum_{b_{2}} p(b_{2}  \vert M_{2})  \rho(b_{2} \vert M_{2})\\&=....=\sum_{b_{n}} p(b_{n} \vert M_{n})\rho(b_{n} \vert M_{n}),
\end{split}
\end{aligned}
\label{op2}
\end{equation}
where $p(b_y \vert M_y)$ and $\rho(b_y \vert M_y)$ respectively represent the probability and the conditional state of the measurer when he performs a measurement $M_y$ with outcome $b_y$.\par
In any experimental proposal or experiment verifying contextuality inequality, it is crucial to close the compatibility loophole, which requires two consecutive preparations (or measurements) to be compatible or co-measurable with each other. This means that two consecutive operations should not have any impact on each other’s results. It’s worth noting that in a communication game model, the closure compatibility is equivalent to the satisfaction of no-signaling condition, because the two consecutive measurements required to be compatible in traditional Kochen-Specker inequality verification are separated in space (not necessarily spatially separated) and carried out by two different parties in the communication game. Therefore, signal transmission is the necessary and sufficient condition for those consecutive operations to be impact-free on each other.
In our case, the preparation no-signalling is defined to be as follows:
\begin{equation}
\begin{aligned}
\sum_{a_x} p(a_x,b_{y}\vert P_x,M_y)=\sum_{a_{x'}} p(a_{x'},b_y\vert P_{x'},M_y),
\end{aligned}
\label{ns1}
\end{equation}
for all preparation events $(a_x|P_x)$ (here an event stands for a preparation and its output) with $a_x\in\{0,1\}$ and $P_x\in\{P_1,P_2,...,P_m\}$. Symmetrically, the measurement no-signalling shall be defined as
\begin{equation}
\begin{aligned}
\sum_{b_y}p(a_x,b_y \vert P_x,M_y)=\sum_{b_{y'}}p(a_x,b_{y'}\vert P_x,M_{y'}),
\end{aligned}
\label{ns2}
\end{equation}
for all measurement events$(b_y|M_y)$ (similarly an event here means a preparation and an input) with $b_y\in\{0,1\}$ and $M_y\in\{M_1,M_2,...,M_n\}$. 
\section{\label{Exp}Experimental setup and data processing}
\begin{figure}[t]
\begin{center}
\includegraphics[width=1\columnwidth]{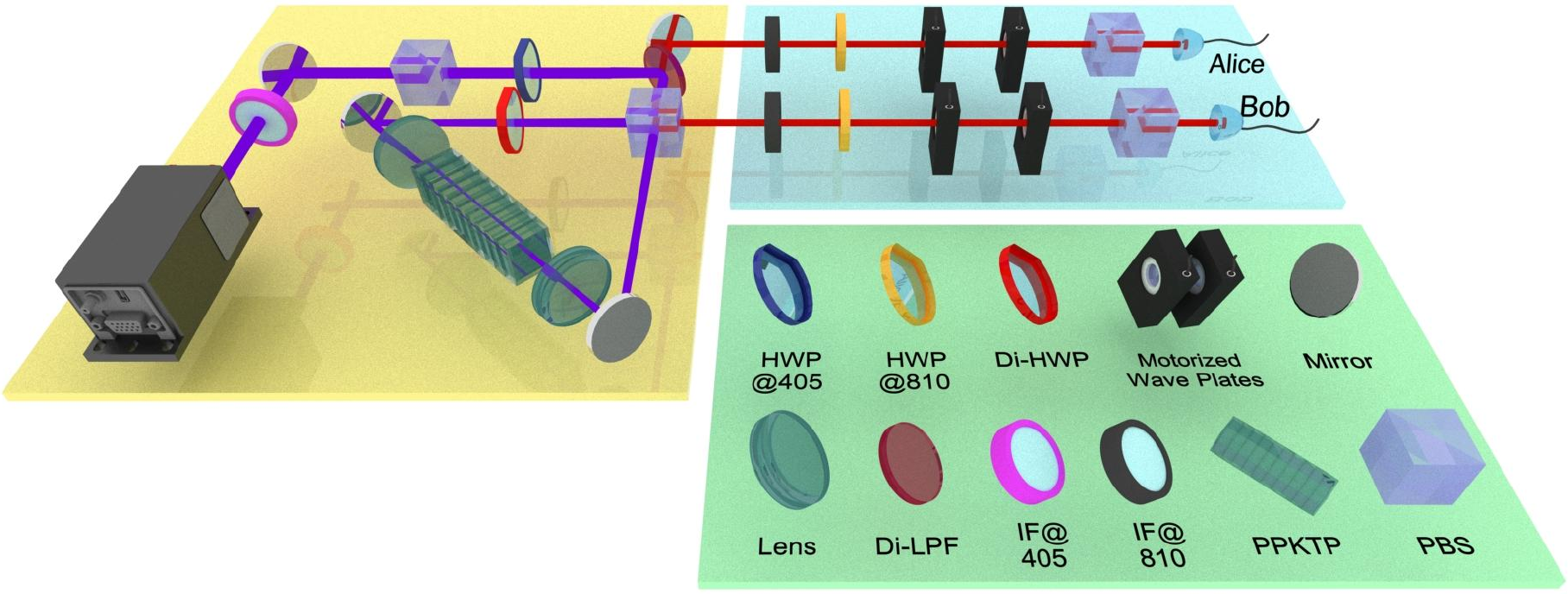}
\caption{Experimental setup. An ultraviolet laser filtered by a narrow-band interference filter with a center wavelength of 405 nm (IF@405) is used to pump a type II periodically poled potassium titanyl phosphate (PPKTP) crystal located in a Sagnac interferometer, to generate a two-qubit polarization-entangled photon state $\vert \psi_{\varphi} \rangle={\rm cos} \varphi \vert HH \rangle +{\rm sin} \varphi \vert VV \rangle$, where $\varphi \in[0, \pi/2]$. After filtering out other stray light by two interference filters with a center wavelength of 810 nm (IF@810), one of the photons is sent to Alice, while the other one is sent to Bob. They will correspondently perform preparation $P_x$ and measurement $M_y$ on the entangled state, by a combination of a quarter wave plate (QWP) and a half wave plate (HWP), both of which are controlled by motorized rotation mount, together with a polarized beam splitter (PBS). The photons are then sent to and detected by the single photon detectors at the end of both sides, and the signals are sent for coincidence.}
\label{setup}
\end{center}
\end{figure}

To experimentally test universal non-trivial contextuality, we use (3,3) and (4,3) communication games as two examples. As shown in Fig.~\ref{setup}, the entangled state is prepared by shining a 405nm laser into a 10mm long periodically polarized potassium titanyl phosphate (PPKTP) crystal. The state we produce can be written in the general form of:
\begin{equation}
\begin{aligned}
\psi_{\rm{AB}}(\theta)=\cos\theta|HH\rangle+\sin\theta|VV\rangle.
\end{aligned}
\label{theta}
\end{equation}
Here the parameter $\theta$ is controlled by rotating the half-wave plate (HWP) after the first polarization beam splitter (PBS), and we take our experiment by preparing the state in a sequence $\theta\in\{0.050, 0.100, 0.152, 0.206, 0.262, 0.322, 0.388, 0.464, 0.560, 0.785\}$.

\begin{figure}[t]
\begin{center}
\includegraphics[width=1\columnwidth]{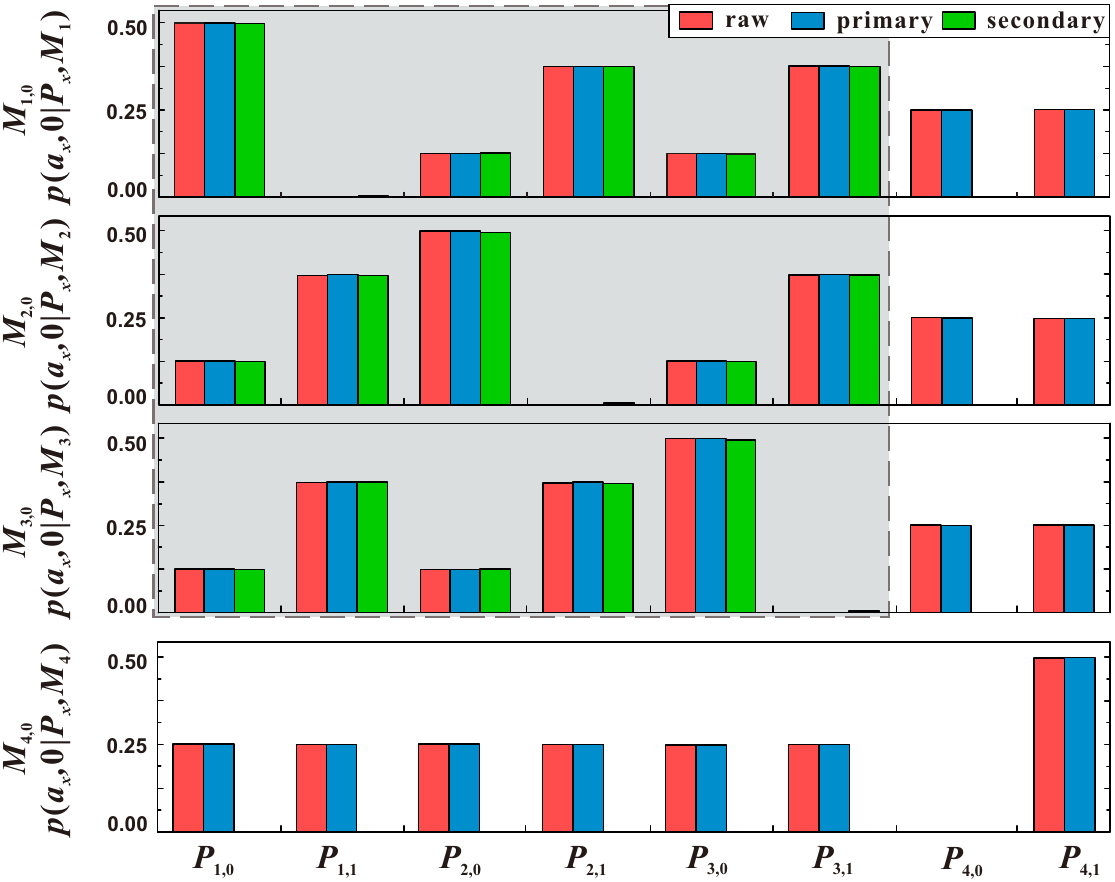}
\caption{\label{outcome}Outcome probabilities for raw, primary, and secondary data. The chart shows outcome probabilities for raw data (in red), primary data (in blue), and secondary data (in green) averaged over 30 runs. For every preparation-measurement pair, we only consider the probability of obtaining outcome 0 in the measurement, as the probability of obtaining outcome 1 is simply the complementary result in statistical analysis. The shaded grey region highlights measurements and preparations for which secondary procedures were found. Error bars are obtained by averaging over 30 runs, which are at most 0.003 and nearly invisible on this scale, as are the discrepancies among raw, primary, and secondary data, which are at most 0.012.}
\end{center}
\end{figure}
After being produced, the entangled pair is sent to test the non-contextuality inequality, the two entangled photons of which are processed and counted respectively on Alice's side for preparation and Bob's side for measurement (it also works by reciprocally exchanging their roles), where the polarization state on each side is handled (named as preparation and measurement respectively) with the composition of a motorized quarter-wave plate (QWP), a motorized HWP and a PBS. For example, in (3,3) scenario, a simple choice of the optimal settings for preparations and measurements would be $P_1 = \sigma_3$, $P_2=\frac{\sqrt3}{2}\sigma_1-\frac{1}{2}\sigma_3$, $P_3=-\frac{\sqrt3}{2}\sigma_1-\frac{1}{2}\sigma_3,$ and $M_1 = -\sigma_3$, $M_2=-\frac{\sqrt3}{2}\sigma_1+\frac{1}{2}\sigma_3$, $M_3=\frac{\sqrt3}{2}\sigma_1+\frac{1}{2}\sigma_3$ \cite{pan2019revealing}. Similarly, while in (4,3) scenario, the optimal directions can be taken as, $P_1 = \frac{1}{\sqrt{3}}(\sigma_1+\sigma_2+\sigma_3)$, $P_2 = \frac{1}{\sqrt{3}}(\sigma_1+\sigma_2-\sigma_3)$, $P_3 = \frac{1}{\sqrt{3}}(\sigma_1-\sigma_2+\sigma_3)$, $P_4 = \frac{1}{\sqrt{3}}(-\sigma_1+\sigma_2+\sigma_3)$, and $M_1 = \sigma_1$, $M_2=-\sigma_2$, $M_3=\sigma_3$, with $\sigma_1$, $\sigma_2$, and $\sigma_3$ standing for the three Pauli operators. Following PBS is the detection part, where photons on both sides are counted by two single photon detectors. Then the counts are further analysed in the coincidence counter section, which will then be taken as the raw probabilities of obtaining each outcome correspondent in real world for every preparation-measurement pair. 
In our experiment, we analyse the data from each run and count the results averaged over 30 runs.\par
\begin{figure}[t]
\begin{center}
\includegraphics[width=1\columnwidth]{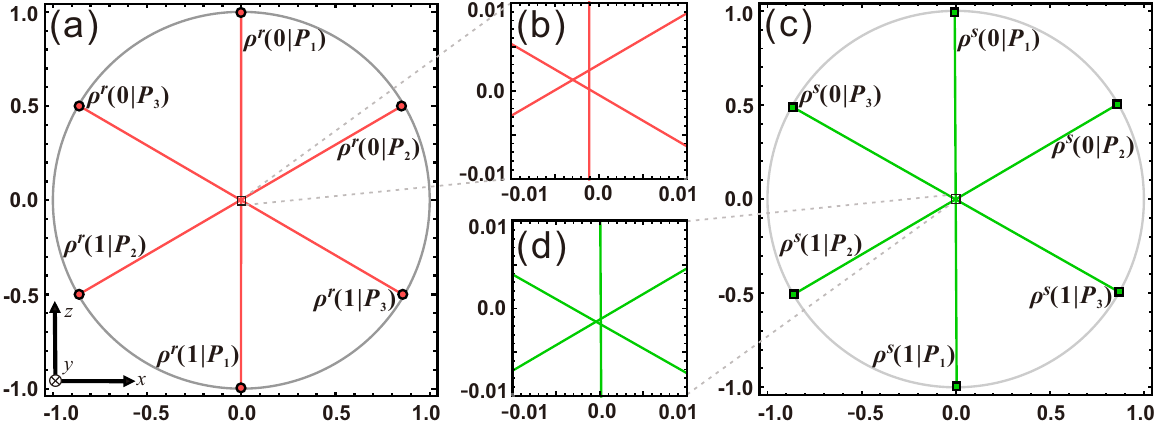}
\caption{\label{operational}Visualization of restored preparation operational equivalence condition in (3,3). (a) Conditional states of the raw preparation procedures (red-colored dots) on the $x-z$ plane of Bloch sphere, where the three red lines connecting each pair of them seemingly intersect at the center. (b) After zooming in by a hundred times, those three red lines actually do not converge, i.e. those preparation procedures are not strictly operationally equivalent with each other. (c) After secondary processing, three green lines converge with each other at one point even on this scale, indicating the restoration of preparation operational equivalence. (d) The correspondent conditional states after secondary preparations procedures (green-colored squares) on the $x-z$ plane of Bloch sphere.}
\end{center}
\end{figure}
To elucidate our data processing part, maximally entangled state $(|HH\rangle+|VV\rangle)/\sqrt{2}$ in (3,3) scenario is shown as an example. We take $P_{x,a_{x}}^t$ as preparation operations with input $x\in\{1, 2, 3\}$ and output $a_{x}\in\{0, 1\}$, and $M_{y,b_{y}}^t$ as measurement operations with input $y\in\{1, 2, 3\}$ and output $b_{y}\in\{0, 1\}$, while $t\in\{r,p,s\}$ denotes how many times of data processing the correspondent operation has experienced, viz. raw data (no processing at all); primary data (after processing only once); and secondary data (after processing twice), which are shown respectively in red, blue, and green in Fig.~\ref{outcome}. 
As is mentioned in the last section, to demonstrate contextuality the lack of operational equivalence needs to be avoided. However, only under unphysical idealizations will the results of raw preparations $P_{x,a_{x}}^r\in$ $\{P_{1,0}^r$, $P_{1,1}^r$, $P_{2,0}^r$, $P_{2,1}^r$, $P_{3,0}^r$, $P_{3,1}^r\}$ and measurements $M_{y,b_{y}}^r\in$ $\{M_{1,0}^r$, $M_{2,0}^r$, $M_{3,0}^r\}$ in realistic experiments strictly match with theoretical predictions, as will be shown later.
In other words, no real experiments can precisely reach ideal operational equivalence, which could only be approximated through mathematical processing. Thus we apply a method that was first mentioned in \cite{mazurek2016experimental} (see Appendix A for more details) for data optimization, which takes two sequential steps to respectively reach tomographic completeness and operational equivalence.\par
In order to impartially evaluate the effectiveness of different models, such as non-contextual or contextual models, we need to utilize generalised probabilistic theories (GPTs) \cite{barrett2007information}, which assume tomographic completeness of three two-outcome measurements, thus having no prejudice upon the underlying model—whether it's quantum, or classical. Thus, the assumption that the system is a qubit is replaced by a weaker assumption that three two-outcome measurements are tomographically complete. Notice that here we need to introduce an extra setting on $\sigma_2$ for both preparation and measurement, i.e. $P_4$ and $M_4$, so as to realize tomographic completeness. Thus, we have the input of raw data as $\{P_{1,0}^r$, $P_{1,1}^r$, $P_{2,0}^r$, $P_{2,1}^r$, $P_{3,0}^r$, $P_{3,1}^r$, $P_{4,0}^r$, $P_{4,1}^r\}$ and $\{M_{1,0}^r$, $M_{2,0}^r$, $M_{3,0}^r$, $M_{4,0}^r\}$, which is fit to a set of states and effects in GPTs for each run, by applying the total weighted least-squares method \cite{krystek2007weighted, press2007numerical}. Our fit returns with a 4$\times$8 matrix to define the primary preparations and measurements, the column of which refers to the $(x,a_x)$ preparation, denoted as $P_{x,a_x}^p$, and the row refers to the $(y,0)$ measurement, denoted as $M_{y, 0}^p$. By taking these generalised states and effects as estimates of primary preparations and measurements, we can achieve tomographic completeness. Fig.~\ref{outcome} shows the results averaged over 30 experimental runs, where raw and primary data are compared but indistinguishable on this scale, the error bars of which are also invisible, fitting well to GPTs.\par
\begin{figure}[t]
\begin{center}
\includegraphics[width=1\columnwidth]{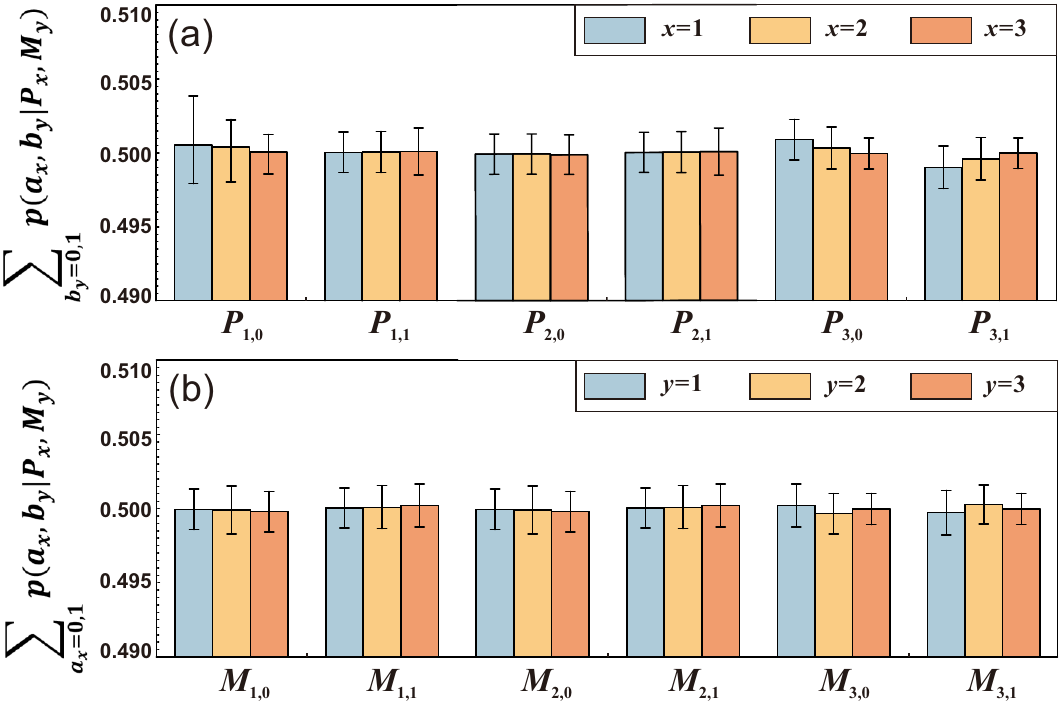}
\caption{Illustration of the closure of compatibility loophole. (a) Different colors refer to the three preparations, while different sets refer to three measurements. As is illustrated in this diagram, all our preparation procedures are the same for three sets of measurements, thus assuring the no-signaling condition from Alice (preparer) to Bob (measurer) by satisfying Eq. (\ref{ns1}), which is satisfied strictly. (b) Different colors refer to the three measurements, while different sets refer to three preparations. Similarly, we assure the no-signaling condition from Bob to Alice by testing Eq. (\ref{ns2}). When both Eqs. (\ref{ns1}) and (\ref{ns2}) are satisfied, the compatibility loophole is closed. It can be seen that the discrepancies between different preparations or measurements are within the error bar, the total average of which is 0.003.}
\label{nosignal}
\end{center}
\end{figure}
However, it should be noticed that in a practical experiment, the primary preparations that are carried out may not satisfy operational equivalence as well, which is shown in Table.~1 in Appendix B. To address this issue, the primary preparations can be processed into "secondary preparations" that are specifically selected to ensure this equivalence. Given the primary statistics, it is possible for us to find within the convex of their mixtures, one set of secondary preparations $P_{1}^s,P_{2}^s,P_{3}^s$ that strictly satisfy equation of preparation operational equivalences (as well as secondary measurements $M_{1, 0}^s,M_{2, 0}^s,M_{3,0}^s$). Within the probabilistic mixtures of this supplemented primary sets $\{P_{1,0}^p, P_{1,1}^p, P_{2,0}^p, P_{2,1}^p, P_{3,0}^p, P_{3,1}^p, P_{4,0}^p, P_{4,1}^p\}$ and $\{M_{1,0}^p, M_{2,0}^p, M_{3,0}^p, M_{4,0}^p\}$, we search for the six secondary preparations $P_{x,a_x}^s = \sum_{x'=1}^4\sum_{a_x'=0}^1u_{x',a_x'}^{x,a_x}P_{x',a_x'}^p$, as well as for the measurements $M_{y,0}^s = \sum_{y'=1}^4v_{y'}^{y}M_{y',0}^p$. Then, by maximizing $C_P = \frac{1}{6}\sum_{x=1}^{3}\sum_{a_x=0}^{1}u_{x,a_x}^{x,a_x}$ over valid $u_{x',a'_x}^{x,a_x}$ while conforming to the constraint of $\frac{1}{2}\sum_{a_x}P_{1,a_x}=\frac{1}{2}\sum_{a_x}P_{2,a_x}=\frac{1}{2}\sum_{a_x}P_{3,a_x}$, we will obtain data which is preparation operationally equivalent. Similarly we maximize $C_M = \frac{1}{3}\sum_{y=1}^{3}v_{y}^{y}$ over valid $v_{y'}^{y}$ to achieve measurement operational equivalence. In Fig.~\ref{outcome}, it is shown that the construction of secondary procedures is also close to the raw data, with discrepancies between them no larger than 0.012, while Fig.~\ref{operational} displays how the outcome probabilities satisfy the operational equivalence. It is clearly demonstrated in Fig.~\ref{operational}(c) that all secondary preparation procedures strictly meet with by converging with each other, thus achieving the operational equivalence condition, which is not satisfied for the original raw data, as is shown in Fig.~\ref{operational}(b).  Note that only preparation operational equivalence of maximally entangled state in (3,3) scenario is taken as an example for exemplification here, while the details of other cases are listed in Table.~1.\par

Then we test whether the compatibility loophole is closed in our experiment, or in other words, whether no-signaling condition is realized. Again, we take the sharing of maximally entangled state in (3,3) scenario as an example. As is shown in Fig.~\ref{nosignal}(a), three preparations represented by different colors (and marked as $x=1$, $x=2$, $x=3$ respectively) are equivalent to each other within a set of measurements, illustrating the satisfaction of Eq. (\ref{ns1}) as preparation no-signaling condition. Similarly, three measurements represented by different colors (and marked as $y=1$, $y=2$, $y=3$ respectively) are equivalent to each other within a set of measurements, illustrating the satisfaction of Eq. (\ref{ns2}) as measurement no-signaling condition. In this way, for both Alice-to-Bob and Bob-to-Alice directions, no-signaling condition is realized within the error of standard deviation, conforming to no-signaling condition as stated before, and indicating the closure of compatibility loophole. For exemplification, our discussion here is focused upon the sharing of maximally entangled state in (3,3) scenario. In Appendix ~\ref{allnosignal} we will quantify how strictly this loophole is closed, for all the states in both scenarios, where we provide detailed results for other six input states, showing that compatibility loophole is also closed in those cases.\par
\section{Results}

\begin{figure}[t]
\begin{center}
\includegraphics[width=1\columnwidth]{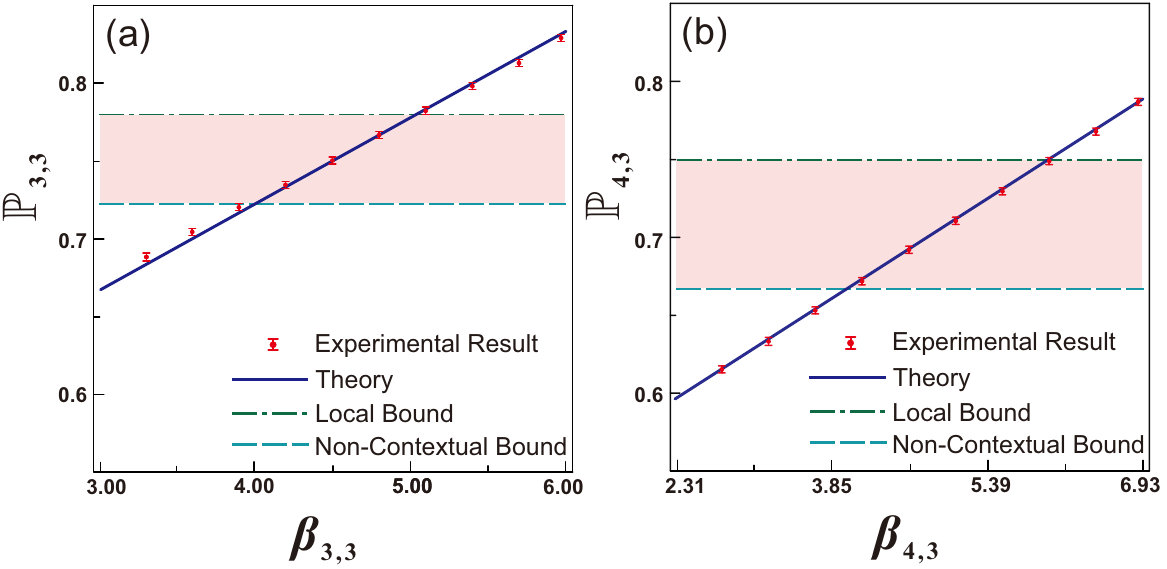}
\caption{\label{result}Experimental results for (3,3) and (4,3) bipartite communication games. (a) In (3,3) scenario, $\mathbb{P}_{3,3}$ is linearly correlated with $\beta_{3,3}$, shown in comparison with the universal non-contextual bound $\mathbb{P}_{3,3}^{UNC} = 0.722$ and $\mathbb{P}_{3,3}^{L} = 0.778$. When the initial state is maximally entangled $ (\vert HH\rangle +\vert VV\rangle)/\sqrt{2}$, the correspondent Bell-like parameter reaches its maximum $\beta_{3,3}=6$, which also achieves the greatest violation of non-contextual inequality $\mathbb{P}_{3,3}=0.832\pm0.001$, exceeding the universal non-contextual bound by 97$\delta$. Error bars in the plots represent the standard deviation in the success probability over the 30 experimental runs. (b)The result of (4,3) scenario also shows a linearly positive correlation between $\beta_{4,3}$ and $\mathbb{P}_{4,3}$, which violates the universal non-contextual bound by 107$\delta$. Error bars in the plots represent standard deviation of the success probability $\mathbb{P}_{4,3}$ over 30 experimental runs.}
\end{center}
\end{figure}

Based on the optimal directions given in the previous section, we can deduce the correspondent maximal success probability of the communication game for a series of entangled states  $\psi_{AB}(\theta)=cos\theta|HH\rangle+\sin\theta|VV\rangle$ in (3,3) scenario:
\begin{equation}
\begin{aligned}
\mathbb{P}_{3,3}^{Q}(\theta)=\frac{4+\sin2\theta}{6}=\frac{1}{2}+\frac{\beta_{3,3}}{18},
\end{aligned}
\label{33}
\end{equation}
where $\beta_{3,3}=3+3\sin2\theta$. For (3,3) scenario, as shown in Fig.~\ref{result}(a), there are two bounds, namely the local bound $\mathbb{P}_{3,3}^{L}\leq14/18\approx 0.778$ (which is equivalent to the assumption of preparation non-contextuality, as noted in \cite{pan2019revealing}) and the universal non-contextual bound $\mathbb{P}_{3,3}^{UNC}\leq13/18\approx 0.722$, which divide the chart into three regions. The left region $\beta_{3,3}\in[3,4)$ shows both Bell locality and universal non-contextuality. The right region $\beta_{3,3}\in[4,6]$, on the other hand, shows quantum non-classicality in both Bell non-locality and universal contextuality, i.e. trivial contextuality. However, the middle region $\beta_{3,3}\in[4,5)$ exhibits simultaneously Bell locality and universal contextuality, also known as non-trivial contextuality. From the discussion above, we find that universal contextuality exceeds Bell non-locality by its lower bound and greater region of violation, which is much easier to satisfy in not only laboratorial demonstration and potential applications. It is clearly shown that experimental results generally align with our theoretical deductions within standard deviation, the largest of which is 0.003.\par

Similarly, in (4,3) scenario we have the correspondent success probability that can also be written in the following form:
\begin{equation}
\begin{aligned}
\mathbb{P}_{4,3}^{Q}(\theta)=\frac{9+\sqrt{3}(1+2\sin2\theta)}{18}=\frac{1}{2}+\frac{\beta_{4,3}}{24},
\end{aligned}
\label{43}
\end{equation}
where $\beta_{4,3}=4\sqrt{3}(1+\sin2\theta)/3$. Fig.~\ref{result}(b) displays two bounds, the local bound $\mathbb{P}_{4,3}^{L}\leq3/4=0.75$ and the universal non-contextual bound $\mathbb{P}_{4,3}^{UNC}\leq2/3\approx 0.666$, and is divided also into three parts: the left $\beta_{4,3}\in[2.31,3.98)$, which displays Bell locality and universal non-contextuality; the middle $\beta_{4,3}\in[3.98,6.00)$, which shows Bell locality but (non-trivial) universal contextuality; and the right $\beta_{4,3}\in[6.00,6.93)$, which demonstrates quantum characteristics in both Bell non-locality and (trivial) universal contextuality. However, (4,3) scenario still differs from (3,3) scenario in several aspects as follows. First, (4,3) scenario has a broader range of non-trivial universal contextuality compared to (3,3) scenario, despite that (3,3) scenario instead has a lower local bound, in which Bell non-locality is much easier to realize. Second, the relationships among different quantum correlations are different in these two scenarios, e.g. the bound of preparation non-contextuality is equivalent to its local bound in (3,3) scenario, while in (4,3) scenario the correspondent bound is equivalent to its universal non-contextual bound. From the figure it can be seen that our result matches with theoretical predictions well, with an averaged error bar of 0.001.

\section{\label{summary}Discussion}
In this work, we provide a generalized bipartite model for the verification of the universal non-contextual inequality, and choose (3,3) and (4,3) scenarios as two examples to test our communication game model. By utilizing the method in Appendix A, our experiment is set free from both operational inequivalence and compatibility loopholes, manifesting a solid step toward the loophole-free demonstration of universal contextuality. We then take a series of states to further investigate the relationships among universal contexutality, Bell non-locality and preparation contextuality. Our result shows that the average success probability of communication game is linearly related to the corresponding Bell-like parameter and solely dependent on the shared state, with the maximum of which realized only when the state is maximally entangled. Notably, Bell non-locality disappears faster than universal contextuality in both scenarios, forming a certain region where quantum theory surpasses the non-contextual bound but not the local bound, and simultaneously revealing Bell locality as well as universal contextuality, also known as non-trivial contextuality. Within this non-trivial region, universal contextuality is more advantageous than Bell non-locality, showing its superiority as a more general quantum resource. It is also worth noting, universal contextuality still outpaces preparation contextuality in (3,3) scenario, despite their convertibility in (4,3) scenario. Our work helps elucidate the nuanced interplay among various quantum correlations, and contributes to deepening the understanding of those fundamental concepts and resources to be applied in future tasks.\par
However, our experiment is not free of all the loopholes in several aspects. One of them is the imperfect detection efficiency, or detection loophole, which has been closed in Bell non-locality and KS contextuality, and is also achievable in the demonstration of universal contextuality either by assuming fair sampling or apply a trapped ion system. Notably, our choice of directions for preparation and measurement may not be optimal for non-maximally entangled states, and further optimization could result in a even greater violation of inequality or a broader region of universal contextuality \cite{horodecki1995violating}. This leaves promising avenues for future research on contextuality, particularly in improving such tests.\par

\section*{Appendices}
\appendix
\section{\label{appendix1}Data Processing Procedures}
In our experiment as well as almost every realistic operational systems, it is almost certain that environmental noises and inaccurate measurement will lead to operational inequivalence. Therefore, a series of further processing of raw data is essential to match with the necessary operational equivalent condition. Here we exemplify how this series of data processing is reached by the following two steps.
First of all, we need to make sure that our experimental results well fit in the framework of generalised probabilistic theories. 
We define $r_{x,a_x}^{y}$ as $p^r(a_x,0|P_{x},M_{y})$, the fraction of outcome 0 returned by measurement $M_{y}$ on preparation $P_{x,a_x}$, the results can be summarized in a 4$\times$8 matrix of raw data, defined as:
\begin{equation}
D^r={
\left( \begin{array}{cccccccc}
r_{1,0}^1 & r_{1,1}^1 & r_{2,0}^1 & r_{2,1}^1 & r_{3,0}^1 & r_{3,1}^1 & r_{4,0}^1 & r_{4,1}^1\\
r_{1,0}^2 & r_{1,1}^2 & r_{2,0}^2 & r_{2,1}^2 & r_{3,0}^2 & r_{3,1}^2 & r_{4,0}^2 & r_{4,1}^2\\
r_{1,0}^3 & r_{1,1}^3 & r_{2,0}^3 & r_{2,1}^3 & r_{3,0}^3 & r_{3,1}^3 & r_{4,0}^3 & r_{4,1}^3\\
r_{1,0}^4 & r_{1,1}^4 & r_{2,0}^4 & r_{2,1}^4 & r_{3,0}^4 & r_{3,1}^4 & r_{4,0}^4 & r_{4,1}^4
\end{array} 
\right)}.
\label{rawdata}
\end{equation}
It is clear that one needs to assume that the measurements one has performed form a tomographically complete set, otherwise statistical equivalence relative to those measurements does not imply statistical equivalence relative to all measurements. Recall that the assumption of preparation non-contextuality only has non-trivial consequences when two preparations are statistically equivalent for all measurements.
And the minimal assumption of which is to ensure that our operation of the four measurements (as well as that of four preparations) are tomographically complete. Note that according to \cite{mazurek2016experimental}, it is proven that four outcome measurements are tomographically complete if and only if $ \alpha p_{x,a_x}^1+\beta p_{x,a_x}^2+\gamma p_{x,a_x}^3+\epsilon p_{x,a_x}^4-1=0$, where $\{\alpha,\beta,\gamma,\epsilon\}\in R$ are real numbers. Thus, to find the GPT-of-best-fit, we simply need to minimize the difference between the pre-processing (raw) and post-processing (primary) versions of data, while satisfying the condition limited by tomographic completeness, which can be paraphrased as a optimization problem:
\begin{equation}
\begin{aligned}
\underset{\left\{p_{x, a_x}^{i}, \alpha, \beta, \gamma, \sigma\right\}}{\operatorname{minimize}} & \chi^{2}=\sum_{x=1}^{4} \sum_{a_x=0}^{1} \chi_{x, a_x}^{2}=\sum_{x=1}^{4} \sum_{a_x=0}^{1} \sum_{y=0}^{4}\frac{(r_{x,a_x}^y-p_{x,a_x}^y)}{\Delta r_{x,a_x}^y}, \\
\text { subject to } & \alpha p_{x, a_x}^{1}+ \beta p_{x, a_x}^{2}+\gamma p_{x, a_x}^{3}+\epsilon p_{x, a_x}^{4}-1=0 \\
& \forall x=1, \ldots, 4, a_x=0,1, y= 1, \ldots, 4.
\end{aligned}
\label{t}
\end{equation}

Solving the problem will return us with a 4$\times$8 optimized matrix of primary data, which can be written as:
\begin{equation}
D^p={
\left( \begin{array}{cccccccc}
p_{1,0}^1 & p_{1,1}^1 & p_{2,0}^1 & p_{2,1}^1 & p_{3,0}^1 & p_{3,1}^1 & p_{4,0}^1 & p_{4,1}^1\\
p_{1,0}^2 & p_{1,1}^2 & p_{2,0}^2 & p_{2,1}^2 & p_{3,0}^2 & p_{3,1}^2 & p_{4,0}^2 & p_{4,1}^2\\
p_{1,0}^3 & p_{1,1}^3 & p_{2,0}^3 & p_{2,1}^3 & p_{3,0}^3 & p_{3,1}^3 & p_{4,0}^3 & p_{4,1}^3\\
p_{1,0}^4 & p_{1,1}^4 & p_{2,0}^4 & p_{2,1}^4 & p_{3,0}^4 & p_{3,1}^4 & p_{4,0}^4 & p_{4,1}^4
\end{array} 
\right)}.
\end{equation}

As for step two, i.e. from primary data to secondary data, we need to introduce the secondary preparations:
\begin{equation}
\begin{aligned}
\mathbf{P}_{x, a_x}^{\mathrm{s}}=\sum_{x^{\prime}=1}^{4} \sum_{a_x^{\prime}=0}^{1} u_{x^{\prime}, a_x^{\prime}}^{x, a_x} \mathbf{P}_{x^{\prime}, a_x^{\prime}}^{\mathrm{p}},
\end{aligned}
\end{equation}
where $u_{x^{\prime}, a_x^{\prime}}^{x, a_x}$ are weights in the mixture.
Similarly, we can construct the secondary measurements by
\begin{equation}
\begin{aligned}
\mathbf{M}_{y,0}^{\mathrm{s}}=\sum_{y^{\prime}=1}^{4} v_{y^{\prime}}^{y} \mathbf{M}_{y^{\prime},0}^{\mathrm{p}}.
\end{aligned}
\end{equation}

\par
To realize an optimized secondary measurement procedures, we need to maximize $C_{\mathrm{P}} \equiv \frac{1}{6} \sum_{x=1}^{3} \sum_{a_x=0}^{1} u_{x, a_x}^{x, a_x}$ for (3,3) scenario (while for (4,3) scenario, $C_{\mathrm{P}} \equiv \frac{1}{8} \sum_{x=1}^{4} \sum_{a_x=0}^{1} u_{x, a_x}^{x, a_x}$ with proper weights $u_{x^{\prime}, a_x^{\prime}}^{x, a_x}$ and $C_{\mathrm{M}} \equiv \frac{1}{3} \sum_{x=1}^{3} v_{y}^{y}$ with proper $v_{y^{\prime}}^{y}$. This maximization is limited by the operational equivalence condition in Eqs.~(\ref{op1}) and (\ref{op2}). The final result is supposed to return a 3$\times$6 matrix in (3,3) scenario, 
\begin{equation}
D_{3,3}^s={
\left( \begin{array}{cccccccc}
s_{1,0}^1 & s_{1,1}^1 & s_{2,0}^1 & s_{2,1}^1 & s_{3,0}^1 & s_{3,1}^1 \\
s_{1,0}^2 & s_{1,1}^2 & s_{2,0}^2 & s_{2,1}^2 & s_{3,0}^2 & s_{3,1}^2 \\
s_{1,0}^3 & s_{1,1}^3 & s_{2,0}^3 & s_{2,1}^3 & s_{3,0}^3 & s_{3,1}^3 \\
\end{array}
\right)},
\end{equation}
and a 3$\times$8 matrix in (4,3) scenario,
\begin{equation}
D_{4,3}^s={
\left( \begin{array}{cccccccc}
s_{1,0}^1 & s_{1,1}^1 & s_{2,0}^1 & s_{2,1}^1 & s_{3,0}^1 & s_{3,1}^1 & s_{4,0}^1 & s_{4,1}^1 \\
s_{1,0}^2 & s_{1,1}^2 & s_{2,0}^2 & s_{2,1}^2 & s_{3,0}^2 & s_{3,1}^2 & s_{4,0}^2 & s_{4,1}^2 \\
s_{1,0}^3 & s_{1,1}^3 & s_{2,0}^3 & s_{2,1}^3 & s_{3,0}^3 & s_{3,1}^3 & s_{4,0}^3 & s_{4,1}^3 \\
\end{array} 
\right)}.
\end{equation}.
\vspace{-0.7cm}
\section{\label{appendix2}Detailed illustration of operational equivalence and no-signaling condition}
We have demonstrated in detail how our result satisfies both operational equivalence and no-signaling condition in Fig.~\ref{operational}, by raising the maximally entangled state in (3,3) scenario as an example. Here we include all those $\theta\in\{$0.000, 0.050, 0.100, 0.152, 0.206, 0.262, 0.322, 0.388, 0.464, 0.560, 0.785$\}$ in both (3,3) and (4,3) scenarios, showing that our experiments are definitely free of those two loopholes.

\subsection{\label{alloperational}The restoration of operational equivalence in all scenarios}
All the preparations $P_{x, a_x}$ and measurements $M_{y, b_y}$ will leave a conditional state within the Bloch sphere (on the Bloch sphere if the state is pure), namely $\rho(a_x|P_x)$ (or $\rho({b_y}|M_y)$), as is shown in Fig.~\ref{operational}. Now, we take the equivalent mass center for a pair of preparation conditional states $\rho(0|P_x)$ and $\rho(1|P_x)$ as $\rho({m|P_x})$ (or take for a pair of measurement conditional states $\rho(0|M_y)$ and $\rho(1|M_y)$ as $\rho({m|M_y})$). Note that trace distance between two states $\rho_1$ and $\rho_2$ is
\begin{equation}
\begin{aligned}
d(\rho_1,\rho_2)=\sqrt{Tr[(\rho_1-\rho_2)^\dagger(\rho_1-\rho_2)]}.
\end{aligned}
\label{Tracediff}
\end{equation}
Thus, by calculating the summed-up trace distances among a set of preparations 
\begin{equation}
\begin{aligned}
\Delta P = \frac{1}{2}\sum_{i\neq j}d(\rho(m|P_i),\rho(m|P_j)),
\end{aligned}
\label{Pop}
\end{equation}
we can quantify how well the preparation operational equivalence is reached. 
Similarly, by summing up trace distances among a set of measurements
\begin{equation}
\begin{aligned}
\Delta M = \frac{1}{2}\sum_{i\neq j}d(\rho(m|M_i),\rho(m|M_j)),
\end{aligned}
\label{Mop}
\end{equation}
the measurement operational equivalence is also quantified. 

The lower this value gets, the smaller the differences for different preparations (or measurements) are, the better operational equivalence is thus realized. In Table.~1 we list those two values for raw ($\Delta P^r$ or $\Delta M^r$), primary ($\Delta P^p$ or $\Delta M^p$) and secondary data ($\Delta P^s$ or $\Delta M^s$) of all different ten states in both (3,3) and (4,3) scenarios. It can be clearly seen that, for all the states in both scenarios, our data processing has significantly lowered $\Delta P$ (or $\Delta M$), despite that the perfectness of operational equivalence is slightly decreased following the entanglement of the state, which is understandable since the average distance for raw data is after all also getting larger itself. In short, we have successfully quantified the the operational equivalence, and restored it by data processing, closing this loophole after two rounds of data processing.\par

\subsection{\label{allnosignal}The closure of compatibility loophole in all scenarios}
Now let's take a glance at the closure of compatibility loophole, or in other word, the satisfaction of no-signaling condition in our experiment. Here we apply secondary data, which performs better than raw data as stated before. Same as the restoration of operational equivalence discussed above, we will sequentially take all the states with $\theta\in\{0.000, 0.050, 0.100, 0.152, 0.206, 0.262,\\ 0.322, 0.388, 0.464, 0.560, 0.785\}$, in both (3,3) and (4,3) scenarios to show that our experiment is in strict alignment with the no-signaling condition. As has been discussed before, in our model the closure of compatibility is equivalent to the no-signaling condition, which requires that there is no signal transmission between two sides, this direction or the other, namely:
Eq.~(5), for from-Alice-to-Bob (preparation) no-signaling condition; and Eq.~(6), for from-Bob-to-Alice (measurement) no-signaling condition. As is exhibited from Table.~2 to Table.~21, different procedures in the same column are equivalent to each other within standard deviation, thus proving the validity of correspondent no-signaling condition and the closure of compatibility loophole.\par

\begin{figure*}[htbp]
\begin{center}
\includegraphics[width=1\columnwidth]{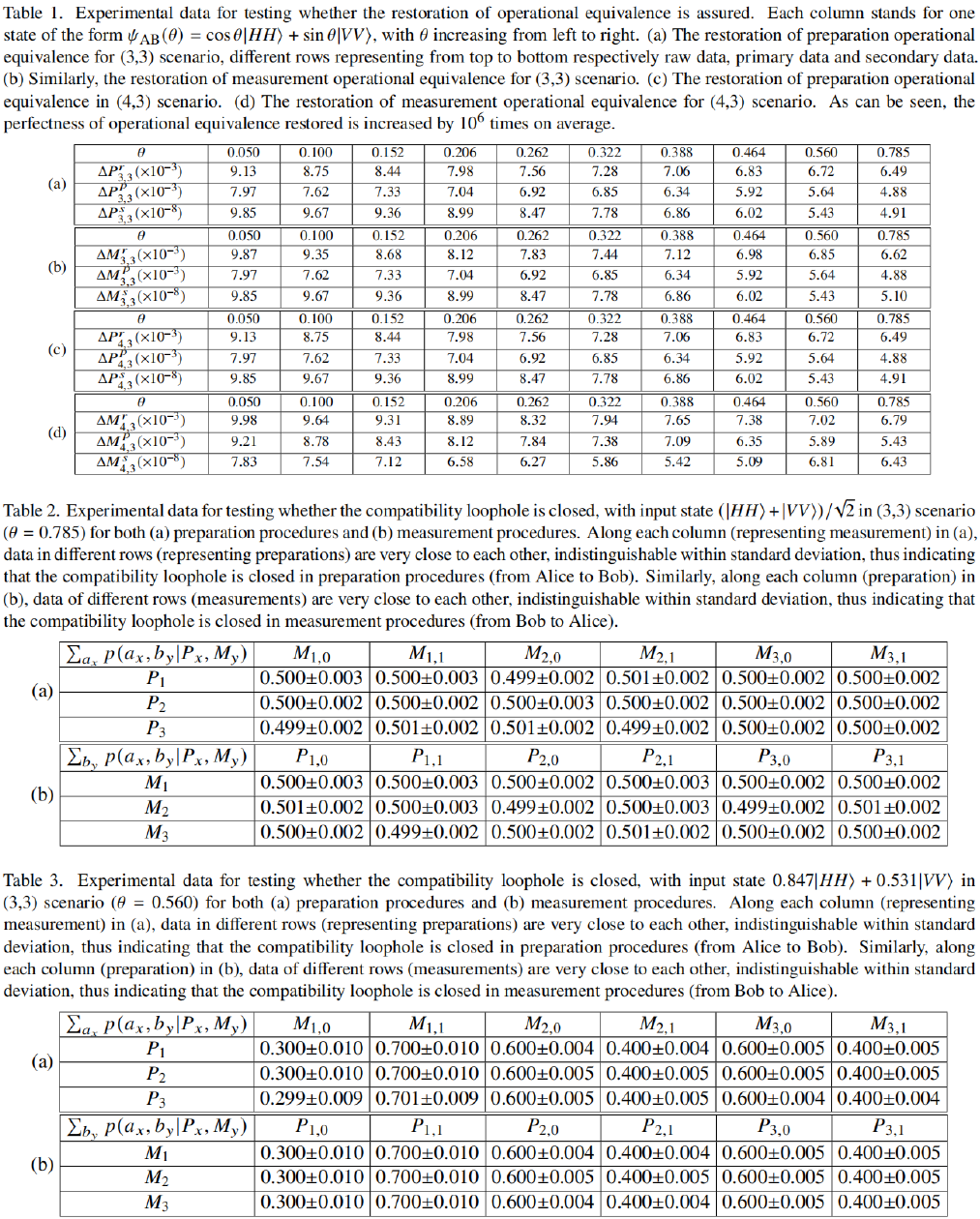}
\end{center}
\end{figure*}
\begin{figure*}[htbp]
\begin{center}
\includegraphics[width=1\columnwidth]{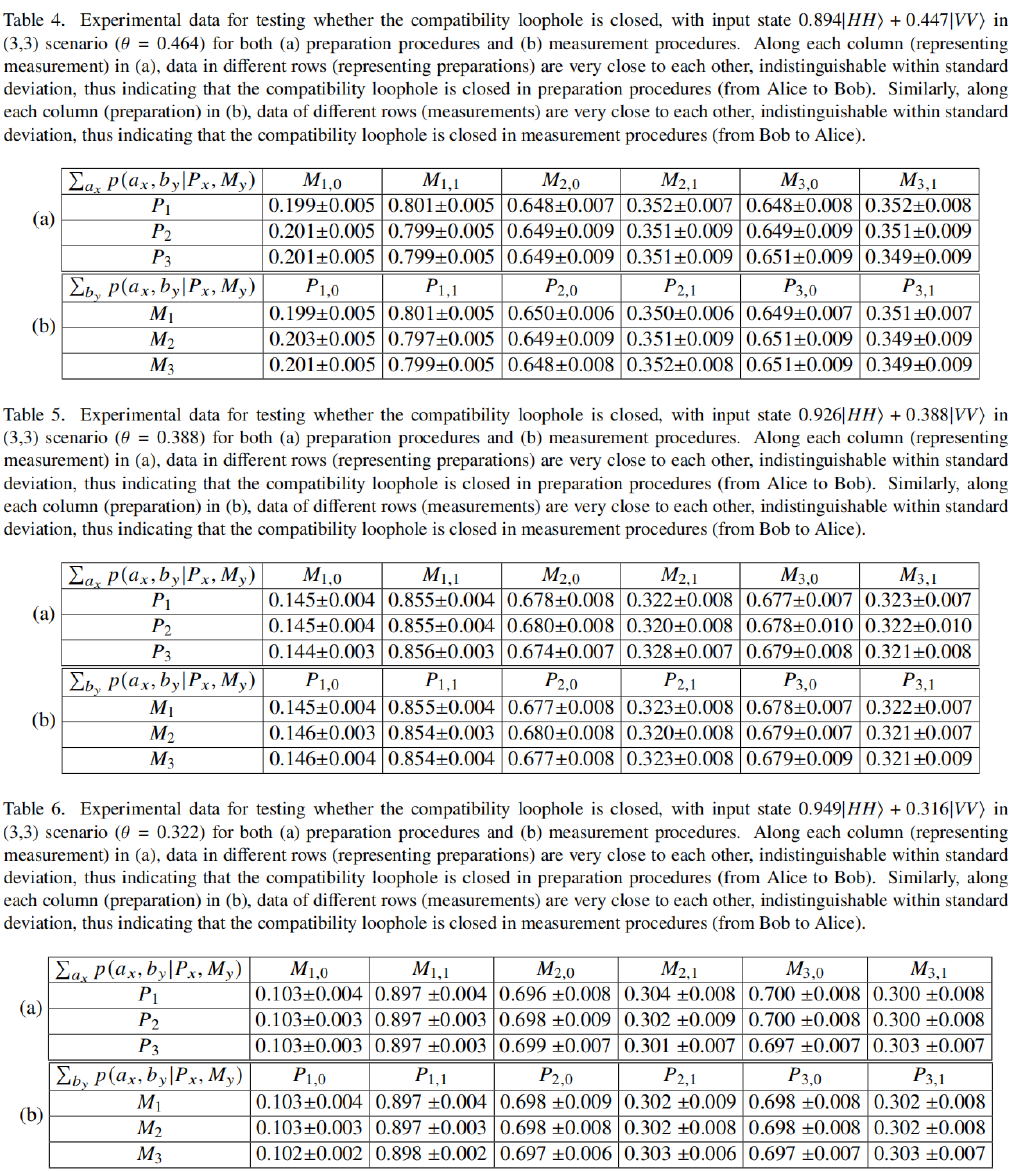}
\end{center}
\end{figure*}
\begin{figure*}[htbp]
\begin{center}
\includegraphics[width=1\columnwidth]{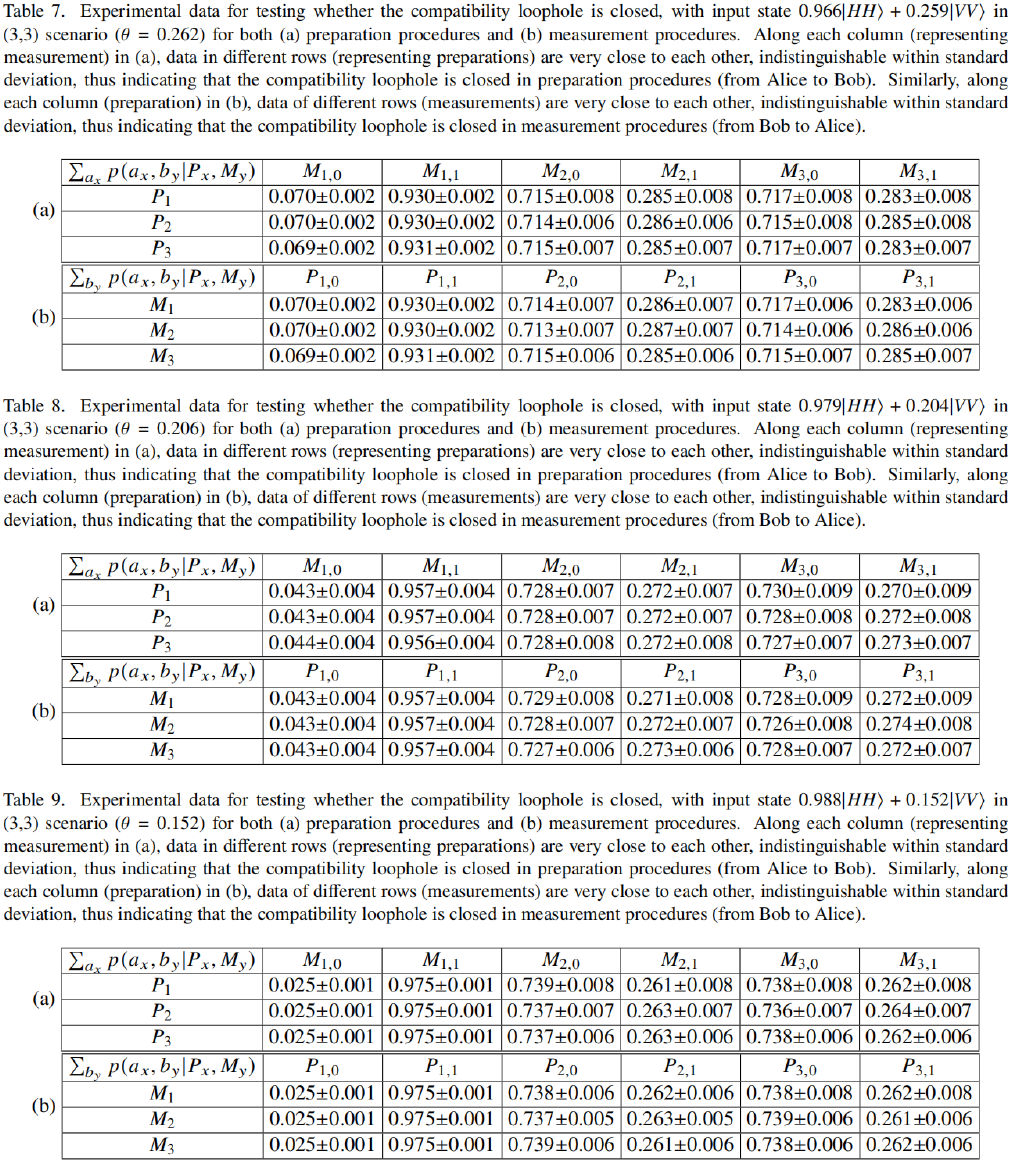}
\end{center}
\end{figure*}
\begin{figure*}[htbp]
\begin{center}
\includegraphics[width=1\columnwidth]{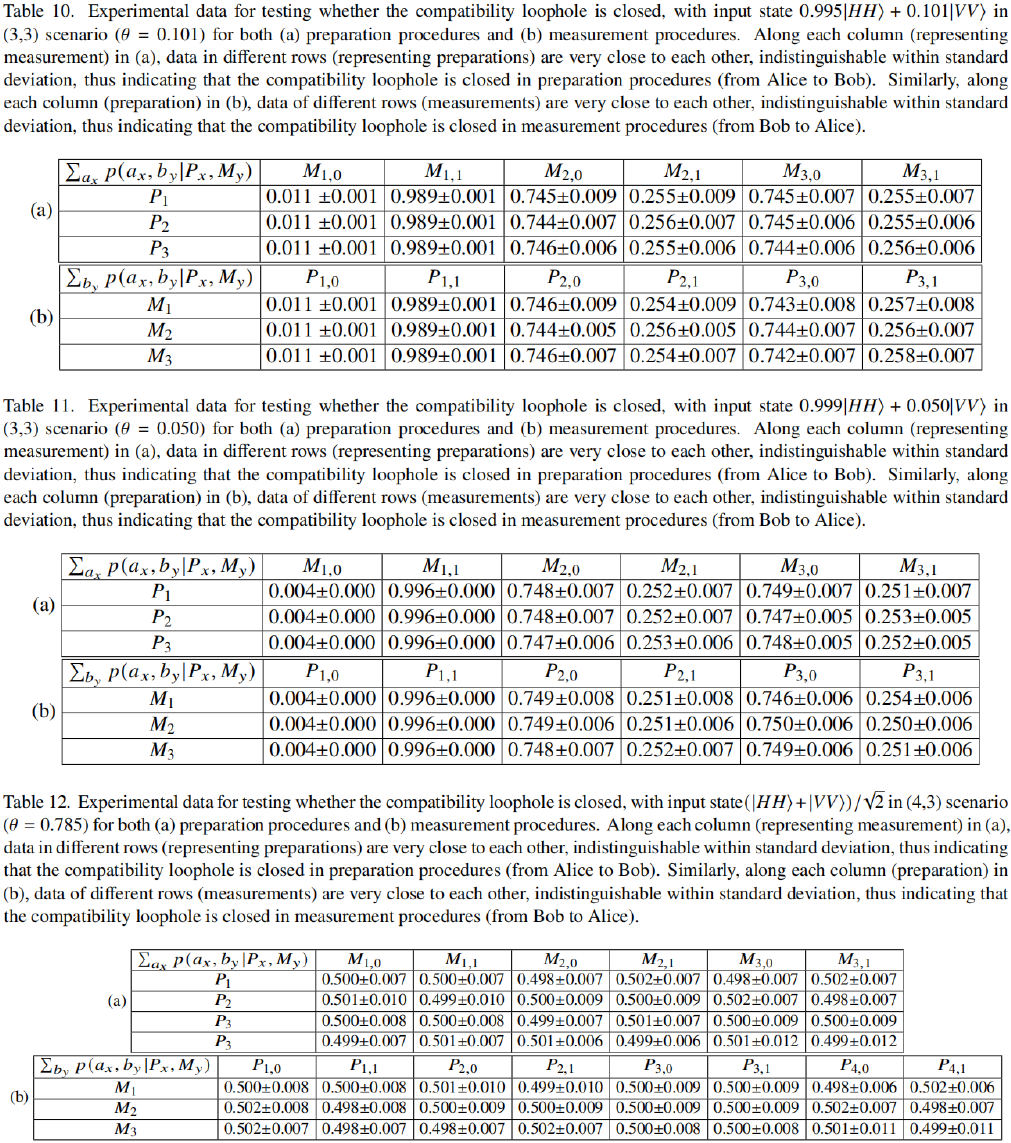}
\end{center}
\end{figure*}
\begin{figure*}[htbp]
\begin{center}
\includegraphics[width=1\columnwidth]{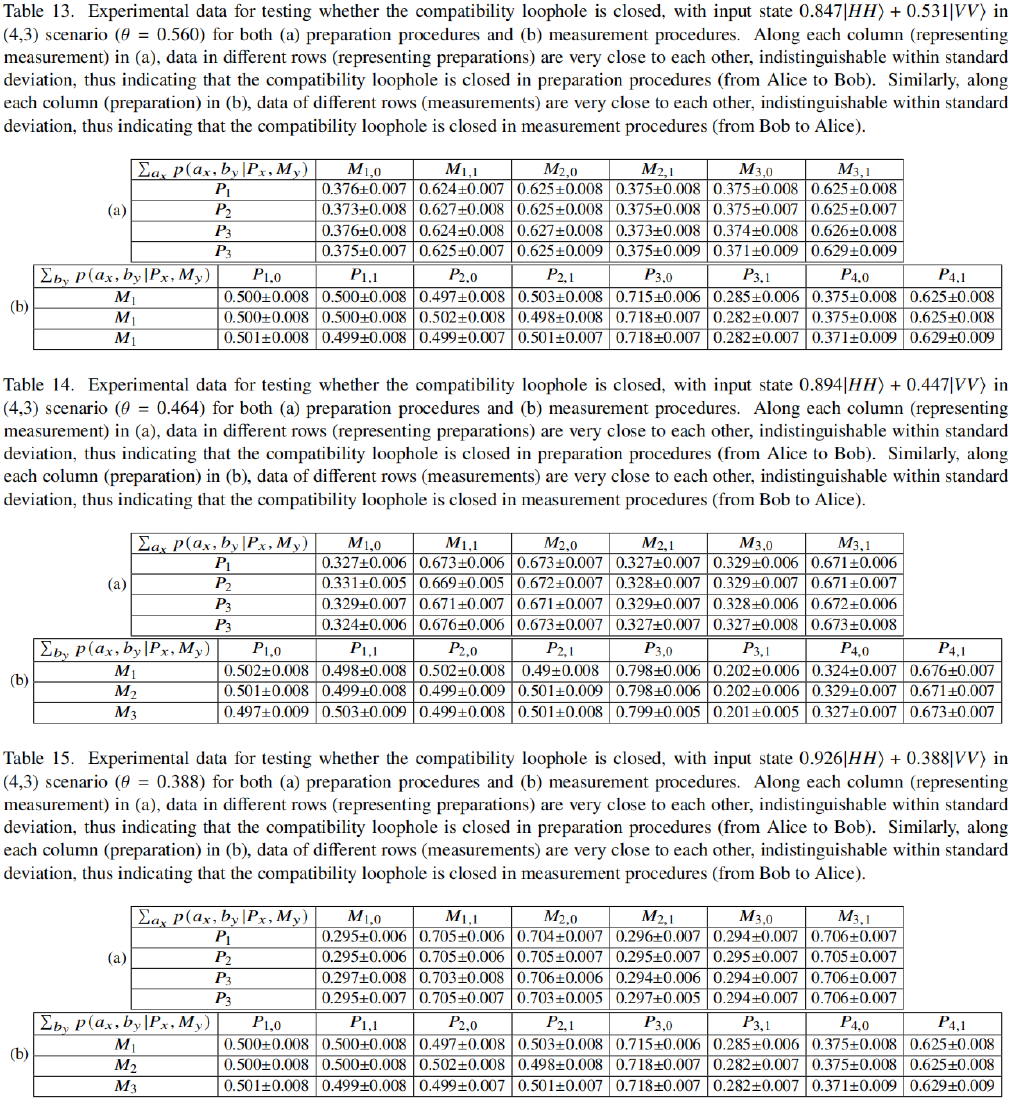}
\end{center}
\end{figure*}

\begin{figure*}[htbp]
\begin{center}
\includegraphics[width=1\columnwidth]{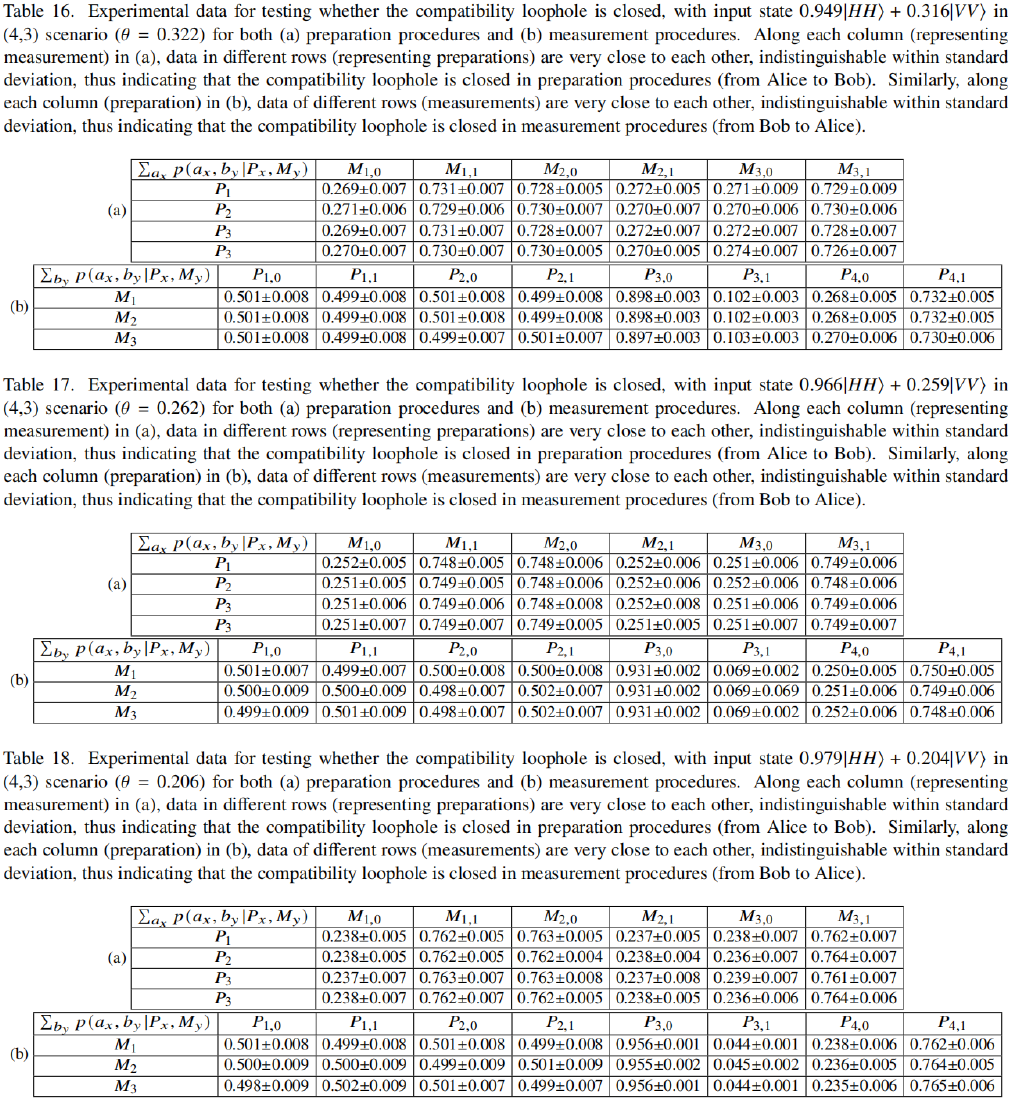}
\end{center}
\end{figure*}

\begin{figure*}[htbp]
\begin{center}
\includegraphics[width=1\columnwidth]{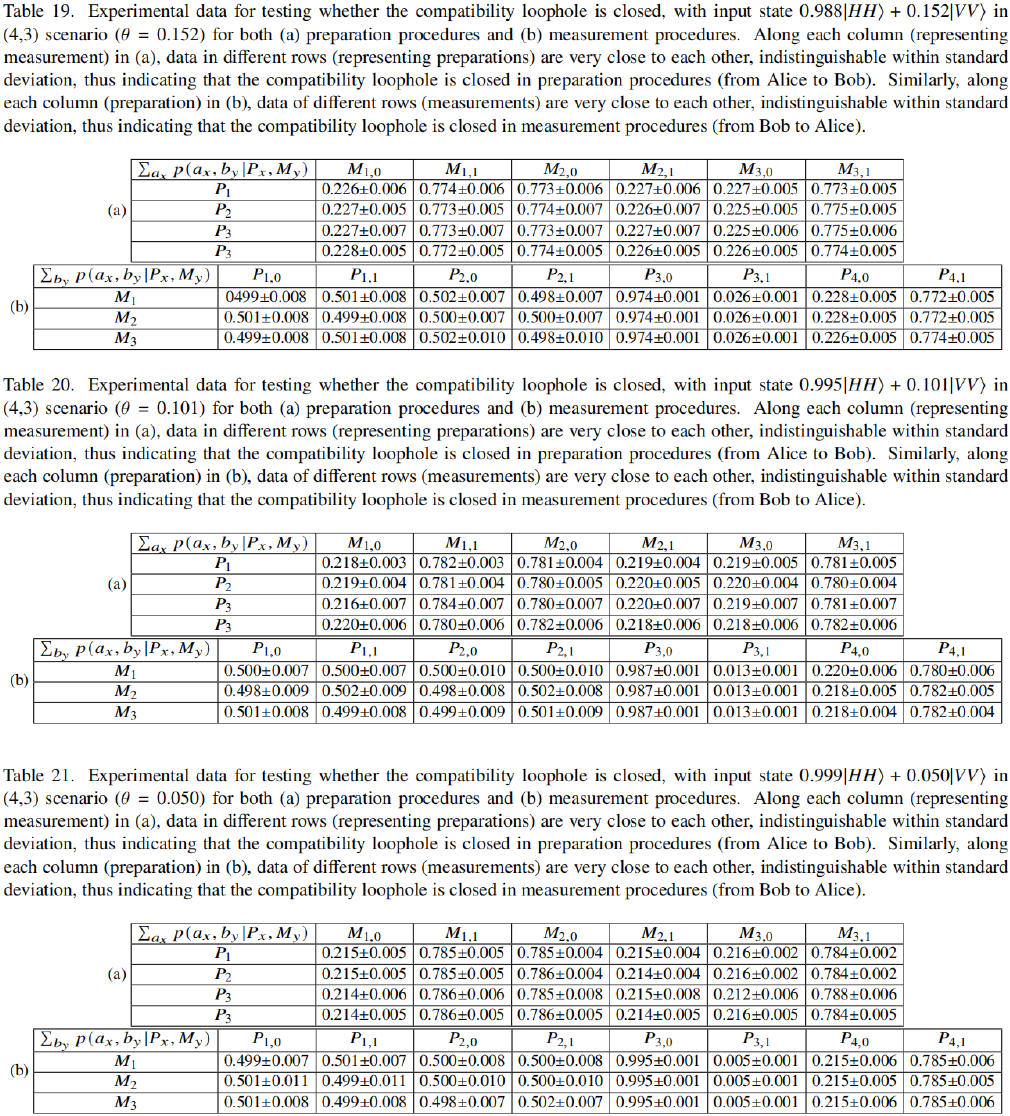}
\end{center}
\end{figure*}
\newpage

\section*{Funding.}
This work was supported by the National Natural Science Foundation of China (Grant No. 12004358), the National Natural Science Foundation Regional Innovation and Development Joint Fund (Grant No. U19A2075), the Fundamental Research Funds for the Central Universities (Grants No. 202041012, No. 841912027 and No. 202364008), the Natural Science Foundation of Shandong Province of China (Grant No. ZR2021ZD19), and the Young Talents Project at Ocean University of China (Grant No. 861901013107).
\section*{Disclosures.}
The authors declare no conflicts of interest.

\section*{Data Availability.}
Data underlying the results presented in this paper are not publicly available at this time but may be obtained from the authors upon reasonable request.

\bibliography{universalbib} 
\end{document}